# A new method for community detection in social networks based on message distribution

Reyhaneh Rigi[a]∗, Mehrdad Jalali[b], Mohammad Hosein Moattar[a]

[a] Department of Software Engineering, Islamic Azad University of Mashhad, Mashhad, Iran.

[b] Institute of Functional Interfaces, Karlsruhe Institute of Technology, Karlsruhe, Germany

**Abstract**

Social networks are the social structures which are composed of people and their relationships and nowadays, play an important role in data extension. In such networks, the communities are recognized as the groups of users who are often interacting with each other. In this article, a method will be introduced for community detection, which has the capability of adoption with different kinds of social networks and also is synchronized with the actual world. One of the most important defined parameters in this paper is the rate of the transferred messages between the nodes of the network, this parameter would be dynamically investigated. In this strategy, the network is reviewed in different time intervals, and the inter-node relations are enhanced or weakened. Therefore, the topology of the network is continuously changing in response to the behavior of the users. The defined parameters in the proposed algorithm are capable of adopting with different types of the social networks and a weight will be assigned to every parameter which is indicative of the relative importance of that parameter in comparison with the other ones. The obtained results show that this method, in comparison with the similar methods, leads to achievement of the desirable results.

## 1. Introduction

Social networks are the new generation of the web sites which today have been in the focus of the users' attention. These types of sites act on the basis of online community formation and each gathers a group of internet users with particular characteristics. Social networks are known as a kind of social media that have provided the access to a new way of communication and content sharing though the internet [1]. Merriam Webster defined the cluster analysis as follows: "a statistical classification method with quantitative comparison of several properties, for discovery of whether the people of the population belong to different groups or not" [2]. The aim of clustering is to sort the samples (people, events and so on) into the clusters in which the members of the clusters are strongly related to each other while the relationship between the members of different clusters is weak [3]. In a social network, the internal relationships of the communities are denser and more complicated than the external relations. The communities provide the valuable data about the type of the users' relationships, the method of data transport between them and the way of the users' distributions in the social networks. In fact, the communities are regarded as the main part of these networks.

∗Corresponding author.
 E-mail addresses : rigi.reyhane@gmail.com (R.Rigi), mehrdad.jalali@kit.edu (M.Jalali), moattar@mshdiau.ac.ir
 (M.H.Moattar).

The community detection in the social networks has played an important role in a broad spectrum of research fields. The main goal of the community detection is that the people in the community have the most similarity with each other. Or in the other words, it can be said that the people in a community are more near to each other and the people of different communities are more far from each other. The fact of being far or near to each other is determined based on different parameters.

In large social networks such as facebook and tweeter, communities can be recognized as groups of users who are often interacting with each other. In the networks, we expect that the amount of the exchanged data between community members be considerably higher than the amount of the exchanged data between community members and the people out of community. The network topology itself expresses whether any two users are connected to each other or not and as a result, the two users can directly transfer messages or not. In fact, it does not provide any signs revealing whether the two users contact to each other or more. In general, we are not aware about the existence of a preferred path along which the data flow [4]. Therefore, the amount of the exchanged data could be applied as a parameter for the purpose of community detection. A considerable amount of the studies on social networks are about the temporal or static networks. But here, the online interactions of the users in different time intervals are under investigation. If a message is transferred between two nodes, a edge is formed between them and by increase of the number of messages, their relationship would be enhanced. If the rate of message exchange increases, the relationship is enhanced otherwise it is weakened.

Next, the related works are reviewed in section 2, and the proposed method would be investigated in section3. In section 4 the evaluation criteria are introduced and the obtained results are presented in section 5. Finally, section 6 is dedicated to conclusion and the future works.

## 2. Background

Numerous community detection algorithms based on the central criteria have been proposed in recent years. For instance, Girvan and Newman, in order to realize the community structure, generalized the vertex Betweenness centricity to the edge Betweenness centricity[5]. This algorithm is one of the most popular hierarchical algorithms. By expansion of Girvan and Newman algorithm, Congo algorithm was introduced for understanding of the interfering communities in the network. This algorithm, similar to the major one, performs the hierarchical clustering but it allows the communities to overlap with each other[6]. Regarding the precision and calculation costs, Louvain algorithm is probably one of the best algorithms [7, 8] and COPRA algorithm is able to find both overlapping and nonoverlapping communities[9]. Oslom algorithm is able to present the higher levels of flexibility, this means that we are allowed to manage the directional and non-directional diagrams to find the overlapping and non-overlapping communities and finally form a hierarchical community[10].

Another strategy for improvement of the community detection algorithms is proposed where a preprocess step, in which the edges are weighted relative to their centers in network topology, is added to the algorithm[4]. In this method, the center of edge is indicative of its contribution in transitive graphs. This preprocess step is added to the three algorithms of Louvain, COPRA and oslom. In ref [11], a new standard is proposed by merging the two concepts of closed walking and clustering coefficients instead of the edge Betweenness known in Girvan and Newman algorithm. In this method, the clustering coefficients are calculated for each of the edges where it is supposed that the edges in the community have the maximum amounts. In this content, by elimination of the edge with the least amounts, it continuously breaks the clustering coefficients into the different clusters.

The authors of ref [12] exclusively analyzed facebook and investigated the friend relationships in this social network, for this purpose, they investigated the topological characteristics of the data indicative diagrams of this online social network by use of two crawling strategies, to be specific BFS and uniform sampling. More analysis on facebook can be found in ref [13].

A new method, called CONCLUDE is proposed in ref [14]. This method of clustering has both the precision and accuracy of the global methods and the scalability of the local ones. In ref [15], first, an interesting model of a social network is introduced. In this model, a link would be formed between each two users in the case of participating in discussion on one or several subjects or stories. So the edges of the network are updating by application of data compatibility between each pairs of users. In second section, a fast parallel modularity optimization algorithm (FPMQA) which does a greedy optimization for discovery of community is applied.

In another paper [16], a community detection method based on the subject is proposed which is a combination of both the clustering of the social subjects and the link analysis. A new hierarchical algorithm for community detection is presented in ref [17]. This approach is based on the density fall between each pairs of the father- son nodes in dendrogram graphs. In this content, higher reduction of density will result in an increase in the probability that the son forms an independent community. Therefore, according to Max-Flow Min-Cut theorem, an algorithm able to automatically find an optimized series from the local communities, has been proposed. An algorithm based on the simple propagation for the community detection which does not require any previous knowledge, has been proposed in ref [18]. The main idea of this method is to present different types of the clusters via a proper method of cluster hierarchical delicate making.

In the scope of graph theory and network analysis, there are different kinds of criteria for vertex centricity in a graph which determine the relative importance of a vertex in a graph. Many of the centricity concepts were first developed in the analysis of the social networks, and many of the expressions for measurement of centricity have been applied as the sociological origin[19]. The most important parameters for centricity are degree, proximity and Betweenness which could be generalized to the edge [20].

The Betweenness centricity is defined as the number of times that a node acts as a bridge in the shortest path between two other nodes, and is introduced by Linton Freeman as a parameter for determination of the quantity of a human control in the relationships between the others in a social network. In a more compact way the Betweenness can be shown as follows:

$$C_B(v) = \sum_{s \neq v \neq t \in V} \frac{\sigma_{st}(v)}{\sigma_{st}} \qquad (1)$$

In 2002, Girvan and Newman presented a definition for the Betweenness centricity of edge which was drastically similar to the one introduced by Anthony. Different margins for Betweenness centricity were proposed by Brands [21]. According to the symbols introduced above, the Betweenness centricity of the edge e ϵ E is defined as follows:

$$C_{B_e}(e) = \sum_{s \neq t \in V} \frac{\sigma_{st}(e)}{\sigma_{st}} \qquad (2)$$

Also, the split Betweenness is defined as: when during the community detection, the Betweenness of a vertex is more than a edge, instead of that edge; the vertex is split [22]. The split Betweenness is applied for support of communities' overlapping.

In the field of social networks, the centricity of a edge helps the strength of social relation manner between two people. The centricity of the edge is useful in expressing the strength quantity of communicational relationships between two things, and therefore it could be helpful in discovery of the new knowledge. In such cases, the virtual communities could be derived from the patterns of interaction between the users, and by their analysis, realize that how a user is able to influence the other one [23]. The method of weighting to the edges has a crucial role in communities' identification, for example, a group of the dense nodes which are connected to each other and have a weak association with a fixed node out of their community, form the communities [24].

### 3. Proposed method

Our proposed algorithm consist of three main step. First the graph topography is updated by application of the rate of the sent messages among the users, secondly identification of the communication edges and eliminating them, finally Community detection. A schematic of the proposed algorithm is shown in fig. 1.

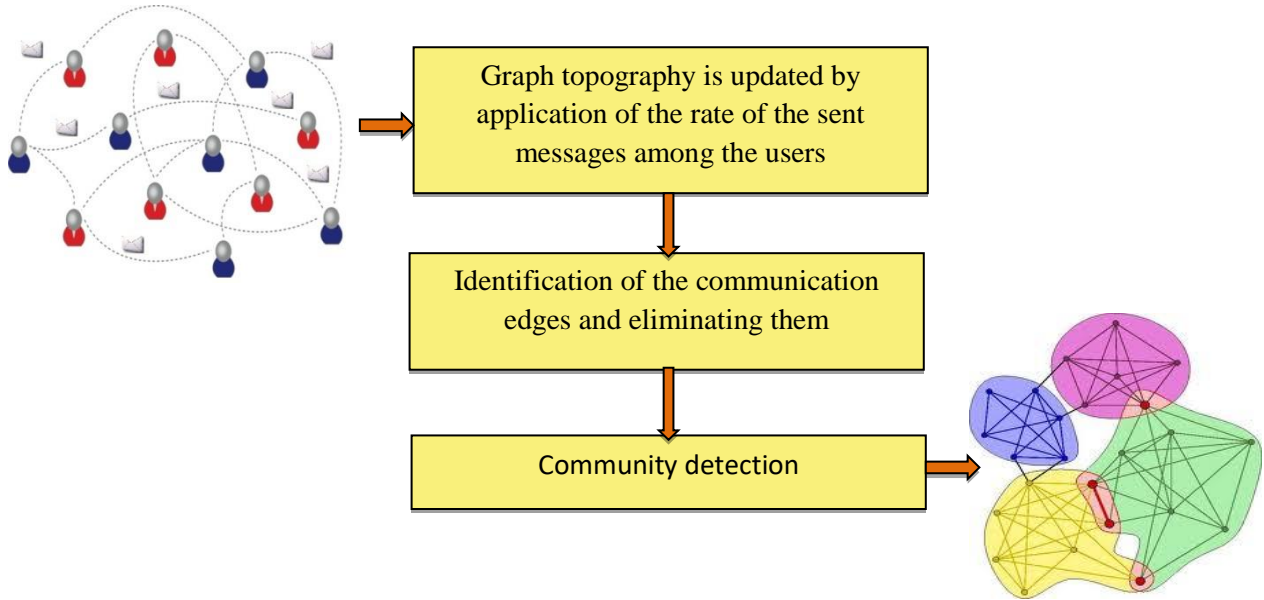

Fig. 1. schematic representation of the proposed algorithm

### 3.1. applied parameters in the proposed algorithms

For the purpose of community detection, the following 3 parameters are implemented in the proposed algorithm:

1) Betweenness centricity of edge
2) Rate of the sent messages (which is investigated dynamically)
3) Similarity of the two nodes.

Since the importance of each parameter is different in each of the social networks, a weight would be assigned to each parameter which is indicator of the relative importance of that parameter in comparison with the other ones.

So a cut parameter would be defined which is expressed as follows:

$$\text{cut}(s, t) = \frac{\alpha_1 (C_B(s,t))}{\alpha_2 (RM(s,t)) \cdot \alpha_3 (S(s,t))} \qquad (3)$$

Where $C_B(s, t)$ is edge Betweenness centricity of s and t nod trial and failure es, $RM(s, t)$ is the rate of the sent messages between S and t nodes and $S(s, t)$ represents the level of similarity between s and t nodes. Three parameters of $\alpha_1$, $\alpha_2$, and $\alpha_3$ are employed for controlling the relative importance of each of the properties. The cut parameter for each node was calculated and then normalized. Afterwards, ranking was performed based on this parameter. The more cut parameter for a edge, it would be regarded as a inter cluster edge.

### 3.2. the proposed algorithm

The proposed algorithm acts as follows:

| **Proposed algorithm** |
|---|
| In each time interval of T:<br>  1- For each node of i and j:<br>    1-1 Calculate the rate of the sent messages in time interval of T between i and j.<br>    1-2 If a message was transferred between i and j in time interval of T, the relationship will be enhanced relative to the number of the messages.<br>    1-3 If there was not any transferred message between i and j in time interval of T, the relationship proportional to the number of the messages will be weakened and if there was not any transferred messages in the previous time intervals, their relating edge will be cut.<br>  2- Calculate cut parameter for each edge and also calculate the split Betweenness for each node.<br>  3- Normalization of the calculated cut and split Betweenness<br>  4- If the maximum value of cut is more than the maximum value of split Betweenness:<br>    4-1- Elimination of a edge with the maximum value of cut<br>    4-2- Recalculation of the cut under the influence of the elimination of edge.<br>  5- Otherwise, if the maximum value of split Betweenness is more than maximum value of cut:<br>   5-1- a node with maximum value of split Betweenness would be divided into two nodes<br>   5-2- recalculation of split Betweenness for each node<br>  6- Repeat from step 3 until the finishing criteria is fulfilled and the communities are formed. |

Fig. 2. the proposed algorithm

### 3.3. Flowchart of the proposed algorithm

Flowchart of the proposed algorithm is shown in fig. 3.

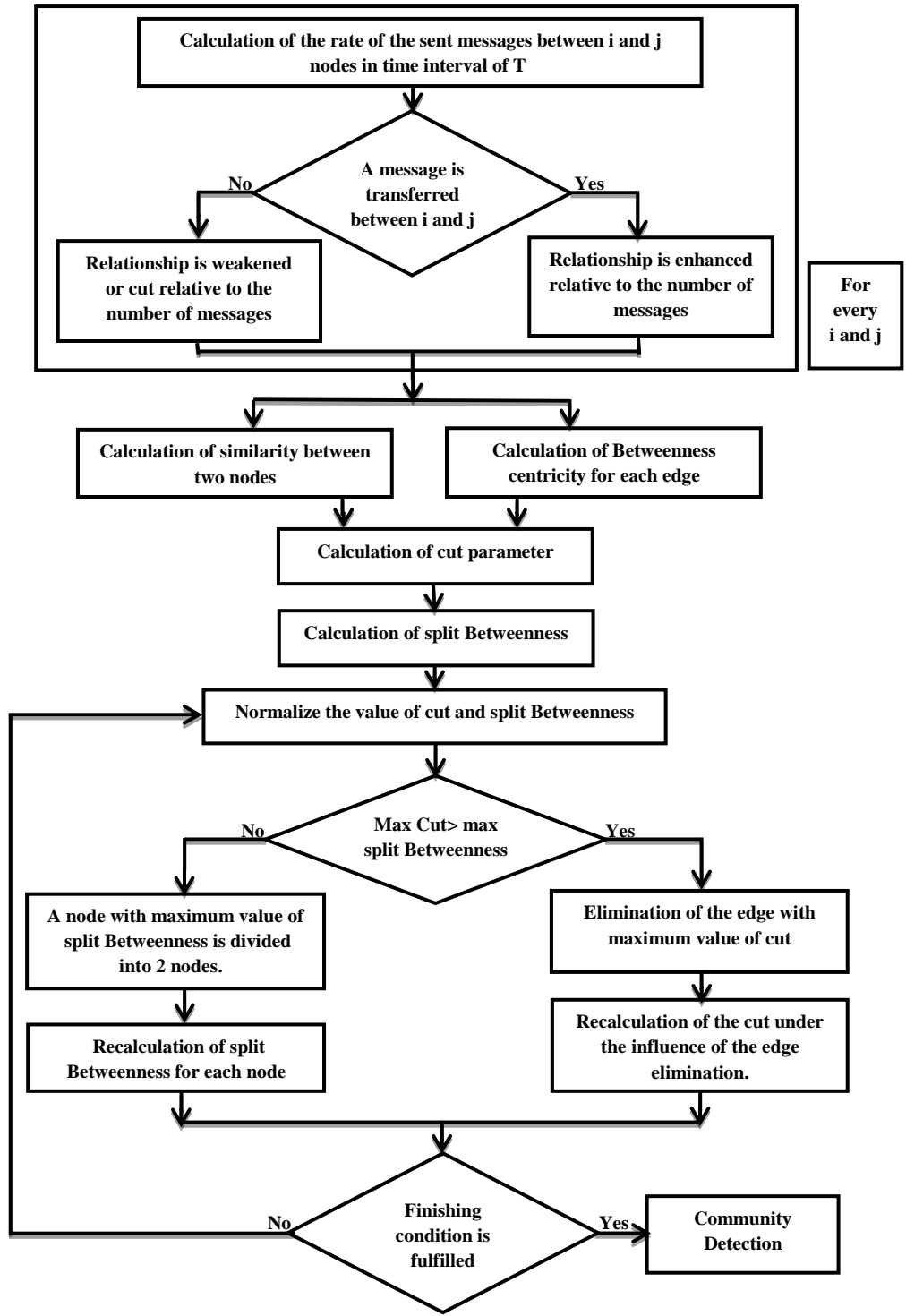

Fig. 3. flowchart of the proposed algorithm

## 3.4. Steps of the algorithm

For the sake of more clarity, the steps of the algorithm are shown in the form of a scenario. In fig. 4.a a graph of a social network with 13 nodes and 26 edges is illustrated.

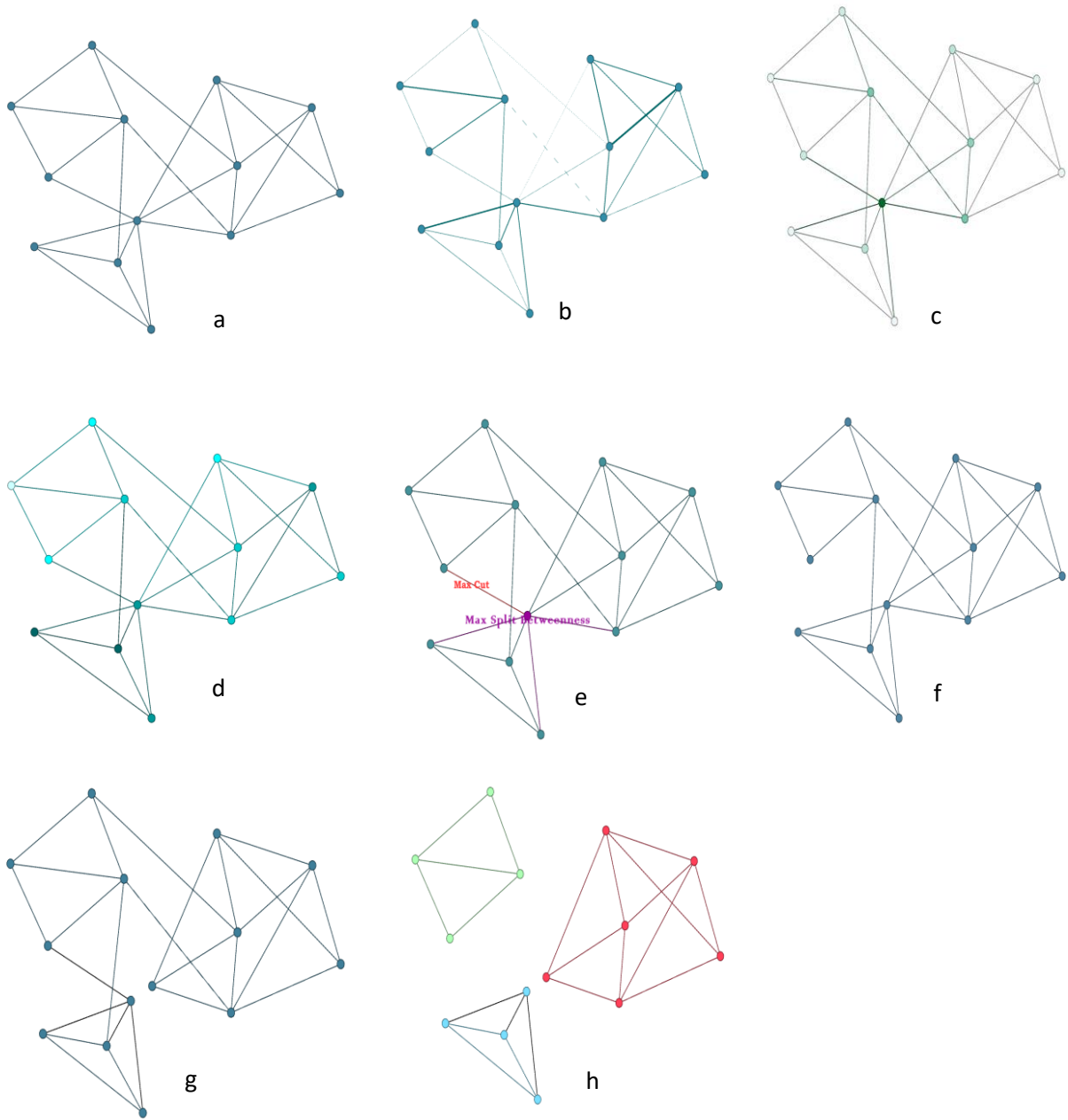

Fig. 4. representation of the steps of algorithm

In the given time intervals, relative to the traffic of the network, the communications of the network are reviewed, if a message is transferred in the mentioned time interval, their communication edge will be enhanced, in a way that this enhancement is proportional to the number of the sent messages between them. If there was not any sent message in that time interval, their relation will be weakened and if there were not any sent messages in next time intervals, their communication edge would be cut. Fig .4.b shows the topology of the graph after these changes. The bolder the communication edge, the more is the rate of the sent messages between them.

Afterwards, the level of betweenness centricity of each edge and the degree of similarity for each two nodes connected to the edge are calculated. Betweenness centricity is depicted in fig .4.c. The bolder the edge, the more betweenness centricity it has. The similarity of the nodes is shown in fig 4.d. The more proximate the color spectrum of two nodes, the more similar they are. In continue, according to the obtained parameters, the parameters of cut and split betweenness would be calculated and then the obtain results would be normalized. In fig .4.e the maximum value of cut for edge and maximum value of split betweenness for the nodes are determined.

If the maximum value of cut is more than the maximum value of the split betweenness, the edge with maximum value will be eliminated which is illustrated in fig .4.f. Otherwise, if the maximum value of split betweenness is more than the maximum value of cut, the node with maximum value of split betweenness will be split into two nodes. In fact this step is applied for performing the overlapping between the nodes. In fig .4.g splitting of a node with maximum value into two nodes is depicted. These steps will be repeated until the communities are formed in the major graphs. After the execution of 5 stages of the algorithms the communities are formed and these obtained communities are shown in fig .4.h.

### 4. Assessment parameters

In this section we describe the Assessment parameters that its usage is beneficial to raise the quality of a community detection algorithm.

#### 4.1. assessment of modularity

In the strategies based on the network modularity, a parameter called modularity is defined which is often shown by Q. this parameter is applied for the quality assessment of Partitioning of a graph G and its aim is to find a partition on G that has the maximum value of Q. the function of the network modularity is defined as follows[4]:

$$Q = \sum_{c=1}^{n_c} \left[ \frac{l_c}{m} - \left(\frac{d_c}{2m}\right)^2 \right] \quad (4)$$

In which $n_c$ is the number of the communities, ،$l_c$ is indicative of the total number of edges connected to vertices in the community c. and $d_c$ is the sum of the degrees of the vertices forming c.

#### 4.2. Quality assessment

In order to make an assessment on the quality of the results, a parameter called normalized mutual information or NMI was applied which is developed by Danon et al in 2005 and had its origins in informatics theory[25]. This assessment is expressed as follows:

$$NMI(A,B) = \frac{-2 \sum_{i=1}^{C_A} \sum_{j=1}^{C_B} N_{ij} \log\left(\frac{N_{ij}N}{N_i \cdot N_j}\right)}{\sum_{i=1}^{C_A} N_i \log\left(\frac{N_i}{N}\right) + \sum_{j=1}^{C_B} N_j \log\left(\frac{N_j}{N}\right)} \quad (5)$$

Where, $N_i$ is indicative of the sum of the elements in i$^{th}$ row in the confusion matrix. If the α algorithm acts completely correct, for each found community j, there is an actual community I exactly in accordance with J.

## 5- experimental results

In this section the experiments conducted for assessment of the algorithm function are described. For performance of the experiments, 5 series of the data whose properties are reported in table 1, were employed.

Table 1: the applied data series

| No | Network | No.nodes | No. message | weighted by | Direccted | Type | Ref |
|----|---------|----------|-------------|-------------|-----------|------|-----|
| 1 | Facebook-like Social Network | 1899 | 59835 | number of characters | yes | online social network | [26] |
| 2 | Facebook-like Social Network | 1899 | 59835 | number of messages | yes | online social network | [26] |
| 3 | Facebook-like Forum Network | 899 | 33720 | number of characters | no | online social network | [27] |
| 4 | Facebook-like Forum Network | 899 | 33720 | number of messages | no | online social network | [27] |
| 5 | Freeman's EIES dataset | 32 | 460 | number of messages | yes | online social network | [28] |

### 5-1 Assessment of the proposed algorithm

For conduction the experiments, first, the data series were divided into 6 phases. In each phase the users data are investigated and according to their interactions, their graph would be updating. If the rate of the sent messages between the users increases their relation is enhanced. Otherwise, the relation decreases. According to the updated graph, the desired parameters are calculated and the proposed algorithm is performed on them. In continue, the steps of users' clustering on the data series of facebook- like forum networks by Gaphi software is depicted.

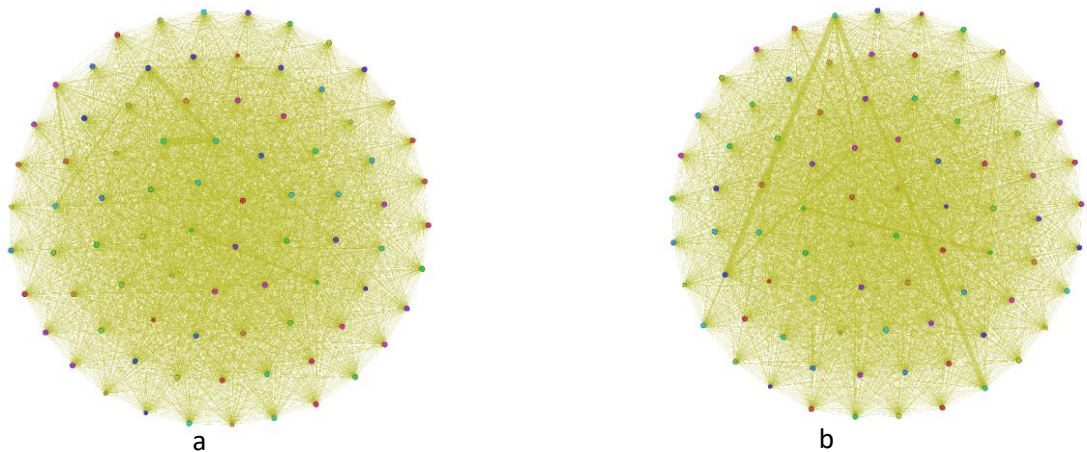

a b

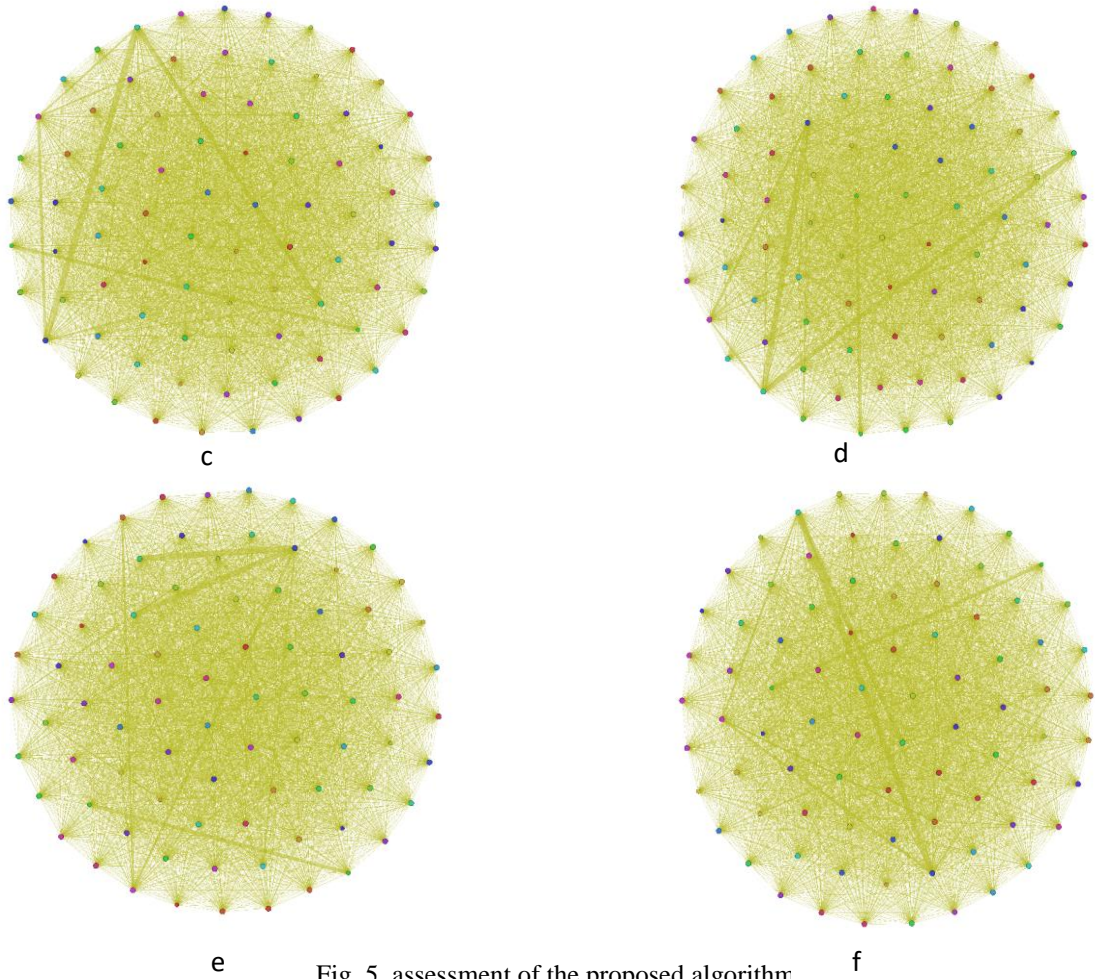

Fig. 5. assessment of the proposed algorithm

## 5-2- assessment of the modularity

In this section, the obtained modularity by each algorithm for each of these data series are calculated. The corresponding results are reported in table 3. By analysis of the table it can be concluded that the use of the proposed algorithm lead to the better results. The functionality of the algorithm, in the case shows a good enhancement.

Table 2: obtained modularity by each algorithm for each of these data series.

| Modularity | proposed algorithm | Closed walks [11] | LM W [4] | CP W [4] | OS W [4] | GN [5] |
|---|---|---|---|---|---|---|
| Facebook-like Social Network (characters) | 0.8539378 | 0.7868432 | 0.7348195 | 0.649756 | 0.5778493 | 0.593719 |
| Facebook-like Social Network (messages) | 0.7564472 | 0.7043612 | 0.7354387 | 0.657623 | 0.6264327 | 0.523622 |
| Facebook-like Forum Network (characters) | 0.8282015 | 0.6834398 | 0.6928638 | 0.737285 | 0.6796725 | 0.614528 |
| Facebook-like Forum Network (messages) | 0.7842003 | 0.7359879 | 0.6152833 | 0.645211 | 0.5903469 | 0.675191 |
| Freeman's EIES dataset | 0.8947311 | 0.7804357 | 0.6954238 | 0.515285 | 0.6180436 | 0.575218 |

From analysis of this table it can be concluded that the use of the proposed algorithm leads to better results. The proposed algorithm in comparison to the presented one in ref [11], showed 8.52% improvement, it was also 12.87% better than Weighted Louvain algorithm[4], and 18.24% improved relative to Weighted COPRA algorithm[4]. Moreover in comparison to Weighted Oslom algorithm[4] it showed about 20.51% improvement. And its enhancement relative to Girvan and Newman algorithm[5] was about 22.70%.

Here, the stages of the algorithm repeat until the best amount for modularity is achieved. In fig .6 the variations of the modularity versus the number of the different repeats of algorithm is shown.

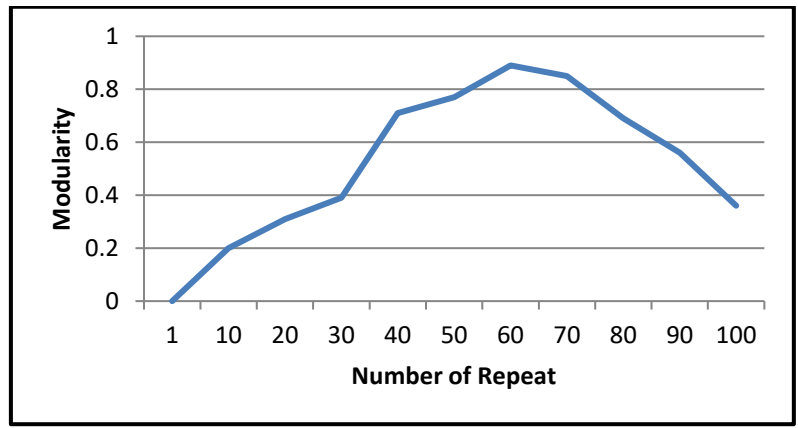

Fig. 6. the variations of the modularity versus the number of the different repeats of algorithm

The best values for $\alpha_1$, $\alpha_2$ and $\alpha_3$ are obtained according to the maximum obtained modularity. We've achieved these three parameters by trial and error. These three parameters are applied for controlling of the relative importance of each properties, and the obtained values are as follows:

Table 3: the best value for $\alpha_1$, $\alpha_2$ and $\alpha_3$ parameters

| | |
|---|---|
| $\alpha_1$ | 0.65 |
| $\alpha_2$ | 0.83 |
| $\alpha_3$ | 0.43 |

### 5-3- assessment of quality

However, calculation of NMI for networks of actual life has still been a challenge since there is no implicit fact in hand for assessment of what communities are in G and what their properties are. Therefore for conducting the experiments, we require to consider the date in which the communities are determined. For this purpose, in this section only the data set of Freeman's EIES dataset is applied. The results are shown in table4.

Table 4: obtained NMI for each algorithm

| algorithms | NMI |
|---|---|
| proposed algorithm | 0.7594185 |
| Closed walks[11] | 0.7182168 |
| LM W[4] | 0.6584127 |
| CP W[4] | 0.5721872 |
| OS W[4] | 0.6032485 |
| GN[5] | 0.5125826 |

The value of NMI ranges from 0 to 1 and the higher values are related to the better algorithms. By analysis of this table it can be concluded that the use of the proposed algorithm leads to better results. The proposed algorithm in comparison to the presented one in ref [11], showed 4.12% improvement, it was also 10.1% better than Weighted Louvain algorithm[4], and 18.72% improved relative to Weighted COPRA algorithm[4]. Moreover in comparison to Weighted Oslom algorithm[4] it showed about 15.61% improvement. And its enhancement relative to Girvan and Newman algorithm[5] was about 24.68%.

## 6- Conclusion and future works

In this paper a new method for communities detection in the social networks is introduced which has the capability of adoption with the different kinds of social networks and is also synchronized with the real world. One of the most important parameters defined in this algorithm is the rate of the transferred messages between the community members which dynamically investigates the network. In this strategy, the network is reviewed in different time intervals and the relationship between the nodes would be enhanced or weakened. Therefore the topology of the network is continuously changing relevant to the behavior of the authors. This method has this property that in addition to protection of the exclusivity of the data, shows the proper actions towards the noise or sabotages in the unsafe environments, and by the least possible amount of data, does the act of community detection. The defined parameters in this algorithm have the capability of adjustment with different social networks they also assign a weight to each parameter which is indicative of the relative importance of that parameter in comparison to the other ones. The propose algorithm tries to resolve or fade the existing challenges in the social networks in a way that it would be able to synchronize with the complications of the real world.

For improvement of the quality of community detection, it is suggested to discover the semantic relationship between the users and also between the virtual networks among them. The hidden relationship between the users is regarded as the virtual union. Discovery of the virtual unions among the users could help in improvement of the community detection based algorithms. Also tagging the contents of the sent messages among the users could considerably help in discovery of the communities.